\documentclass[onecolumn,10pt,showpacs,
amsmath,amssymb]{revtex4}
\usepackage{graphicx} 
\usepackage{hyperref} 
\usepackage{amssymb, amsmath}
\usepackage{epsf}

\def\la{\; \raise0.3ex\hbox{$<$\kern-0.75em\raise-1.1ex\hbox{$\sim$}}\;}
\def\ga{\;  \raise0.3ex\hbox{$>$\kern-0.75em\raise-1.1ex\hbox{$\sim$}}\;}

\def\la{\; \raise0.3ex\hbox{$<$\kern-0.75em\raise-1.1ex\hbox{$\sim$}}\;}
\def\ga{\;  \raise0.3ex\hbox{$>$\kern-0.75em\raise-1.1ex\hbox{$\sim$}}\;}
\def\pFn{p_{\raise-0.3ex\hbox{{\scriptsize F$\!$\raise-0.03ex\hbox{\rm n}}}}
}  

\def\pFa{p_{\raise-0.3ex\hbox{{\scriptsize F$\!$\raise-0.03ex\hbox{$i$}}}}
}  

\def\pFas{p_{\raise-0.3ex\hbox{{\scriptsize F$\!$\raise-0.03ex\hbox{$k$}}}}
}  

\def\pFb{p_{\raise-0.3ex\hbox{{\scriptsize F$\!$\raise-0.03ex\hbox{$\beta$}}}}
}  

\def\vFa{v_{\raise-0.3ex\hbox{{\scriptsize F$\!$\raise-0.03ex\hbox{$i$}}}}
}  
\def\pFp{p_{\raise-0.3ex\hbox{{\scriptsize F$\!$\raise-0.03ex\hbox{\rm p}}}}
}  
\def\pFe{p_{\raise-0.3ex\hbox{{\scriptsize F$\!$\raise-0.03ex\hbox{\rm e}}}}
}  
\def\pFmu{p_{\raise-0.3ex\hbox{{\scriptsize F$\!$\raise-0.03ex\hbox{\rm
$\mu$}}}} }  
\def\m@th{\mathsurround=0pt }
\def\eqalign#1{\null\,\vcenter{\openup1\jot \m@th
   \ialign{\strut$\displaystyle{##}$&$\displaystyle{{}##}$\hfil
   \crcr#1\crcr}}\,}

\newcommand{\vp}{\mbox{${\pmb p}$}}         
\newcommand{\vps}{\mbox{${\vp '}$}}         
\newcommand{\vQ}{\mbox{${\pmb Q}$}}         
\newcommand{\vQa}{\mbox{$\vQ_{i}$}}         

\newcommand{\api}{\mbox{$a_{\pmb p}^{(i)}$}}
\newcommand{\apc}{\mbox{$a_{\pmb p}^{(i) \dagger}$}}

\newcommand{\aps}{\mbox{$a_{{\pmb p} '}^{(k)}$}}
\newcommand{\apsc}{\mbox{$a_{{\pmb p} '}^{(k) \dagger}$}}

\newcommand{\thp}{\mbox{$\theta_{\pmb p}^{(i)}$}}
\newcommand{\thpQ}{\mbox{$\theta_{{\pmb p}+{\pmb Q}_{i}}^{(i)}$}}
\newcommand{\thps}{\mbox{$\theta_{{\pmb p} '}^{(k)}$}}
\newcommand{\thpsQ}{\mbox{$\theta_{{\pmb p} ' +{\pmb 
Q}_{k}}^{(k)}$}}
\newcommand{\uup}{\mbox{$u_{\pmb p}^{(i)}$}}
\newcommand{\vvp}{\mbox{$v_{\pmb p}^{(i)}$}}

\newcommand{\fp}{\mbox{$\mathfrak{f}_{\pmb p}^{(i)}$}}
\newcommand{\np}{\mbox{$n_{\pmb p}^{(i)}$}}

\newcommand{\mN}{\mbox{$\mathcal N$}}
\newcommand{\mF}{\mbox{$\mathcal F$}}
\newcommand{\dd}{\mbox{d}}                     

\begin{document}

\title{The relativistic entrainment matrix of a superfluid nucleon-hyperon 
mixture. II. Effect of finite temperatures}
%
\author{Mikhail E. Gusakov$^{1}$, Elena M. Kantor$^{1}$, Pawel Haensel$^{2}$}
\affiliation{
$^{1}$ Ioffe Physical Technical Institute,
Politekhnicheskaya 26, 194021 Saint-Petersburg, Russia
\\
$^{2}$ N. Copernicus Astronomical Center,
Bartycka 18, 00-716 Warsaw, Poland
}
\date{}
%


\begin{abstract}
We calculate the important quantity of superfluid hydrodynamics, 
the relativistic entrainment matrix
for a nucleon-hyperon mixture
at {\it arbitrary} temperature.
In the nonrelativistic limit this matrix is also termed the
Andreev-Bashkin or mass-density matrix.
Our results can be useful for modeling 
the pulsations of massive neutron stars
with superfluid nucleon-hyperon cores 
and for studies of the kinetic properties 
of superfluid baryon matter.
\end{abstract}

\maketitle

\section{Introduction}
\label{1}

It is generally accepted that observations of 
pulsating neutron stars may potentially provide unique information 
on properties of superdense matter.
For a correct interpretation of the observations one has 
to develop a hydrodynamic theory describing global pulsations. 

In this paper we mainly focus 
on the hydrodynamics of superfluid nucleon-hyperon matter 
of massive neutron-star cores.
More specifically, we study the important 
set of quantities of such hydrodynamics, 
the relativistic entrainment matrix.
We assume the following matter composition:
electrons (e), muons ($\mu$), neutrons (n), protons (p), 
$\Lambda$-, and $\Sigma^-$-hyperons ($\Lambda$ and $\Sigma$, respectively).
According to majority of calculations, 
$\Lambda$ and $\Sigma^-$ are the hyperons, 
which appear first in the stellar matter 
with the increasing density 
(see, however, Ref.\ \cite{fg07} 
and the discussion in Sec.\ I of Ref.\ \cite{gkh09}).

Let us explain the physical meaning 
of the relativistic entrainment matrix.
It is well known, that any baryon species 
$i$ ($i=n$, $p$, $\Lambda$, or $\Sigma$) 
becomes superfluid when temperature $T$ falls below 
some critical value $T_{ci}$. 
The microscopic calculations 
\cite{bb98,ykgh01,ls01,yls99,lp04,vt04,pgw06,tnyt06} 
predict that protons and hyperons 
pair in the spin singlet ($^1$S$_0$) state,
while neutrons in the core pair 
in the spin triplet ($^3$P$_2$) state.
In the most general case, when all baryon species are superfluid, 
we have four condensates 
and a normal (nonsuperfluid) liquid component.
The normal component includes electrons, muons, 
and baryon Bogoliubov excitations. 
We assume that, because of collisions, 
all these `normal' particles have the same hydrodynamic velocity.
Below in this paper we always work in the frame, 
{\it comoving} with the normal liquid component.

If there are no superfluid currents in the system, 
then Cooper pairs are formed by particles 
with strictly opposite momenta 
(e.g., ${\pmb p}$ and $-{\pmb p}$). 
The presence of superfluid currents 
corresponds to a situation
when pairing occurs between particles 
with momenta 
(${\pmb p}+{\pmb Q}_i$) and ($-{\pmb p}+{\pmb Q}_i$).
The total momentum of a Cooper pair 
is then $2 {\pmb Q}_i$.
In this case the particle current density 
${\pmb j}_i$ of a species $i$
can be written as \cite{ga06,gkh09} 
%
\begin{equation}
{\pmb j}_i = 
c^2 \sum_k \, Y_{ik} \, {\pmb Q}_k.
\label{jnn}
\end{equation}
Here the summation is performed over all baryon species, 
$k=n$, $p$, $\Lambda$, and $\Sigma$;
$c$ is the speed of light.
Finally, $Y_{ik}$ is the $4 \times 4$ relativistic entrainment matrix. 
It is symmetric, $Y_{ik}=Y_{ki}$, 
and generally depends on the four baryon number densities $n_i$ 
and temperature $T$.
It follows from Eq.\ (\ref{jnn}) that the superfluid motion of one species
contributes to particle current density of another species (and vice versa).
This `entrainment' effect was first suggested 
by Andreev and Bashkin \cite{ab75}
in the context of superfluid solutions of $^3$He in $^4$He.

In the nonrelativistic limit the matrix $Y_{ik}$ 
(more precisely, its analog, the matrix $\rho_{ik}$)
was analyzed for a neutron-proton mixture 
at zero temperature in Refs.\ \cite{bjk96,cj03,ch06} 
and at arbitrary temperature in Ref.\ \cite{gh05}.
In our recent paper \cite{gkh09} (hereafter GKH09)
we calculate the relativistic entrainment matrix $Y_{ik}$ 
for the nucleon-hyperon matter in the approximation of zero temperature.
The aim of the present study is to extend 
the analysis of GKH09 to finite temperatures.
For that, the strongly interacting nucleon-hyperon mixture is
considered in the frame of {\it relativistic} Landau theory of Fermi liquids, 
generalized to allow for possible superfluidity of baryons.

In Eq.\ (\ref{jnn}) it is assumed that 
the matrix elements $Y_{ik}$ are scalars.
Strictly speaking, this is 
the case
only if
the matter is isotropic
in the absence of superfluid currents,
i.e. if all baryon species pair in the spin singlet ($^1$S$_0$) state.
Meanwhile, neutrons undergo triplet-state pairing.  
This leads to anisotropic neutron energy gap 
$\Delta^{(n)}_{\pmb p}$
and to appearance of the preferred direction 
along the neutron quantization axis.
As a consequence, the matrix elements $Y_{ik}$ 
are tensors with respect to spatial rotations 
(rather than scalars, as in the isotropic case).
This makes the problem of calculation of the matrix $Y_{ik}$
much more complex and model-dependent \cite{gh05}.

To avoid this difficulty we assume, 
following Refs.\ \cite{bhy01,gh05},
that the unperturbed matter can be treated as a collection
of microscopic domains with 
chaotically directed neutron quantization axis.
After averaging over the large number of such domains, 
the elements $Y_{ik}$ 
will become scalars. 
One can obtain then qualitatively correct results 
for the averaged matrix elements $Y_{ik}$, 
assuming $^1$S$_0$ pairing of neutrons,
and introducing an {\it effective} isotropic energy gap 
$\Delta^{(n)}_{\rm eff}(T)$
in the neutron dispersion relation,
\begin{equation}
\Delta^{(n)}_{\rm eff}(T)= \min
\left\{ \Delta^{(n)}(|{\pmb p}|=p_{{\rm F}n}) \right\}.
\label{effGAP}
\end{equation}
Here $\Delta^{(n)}_{\rm eff}(T)$
is defined as the minimum of the angle-dependent
gap $\Delta^{(n)}_{\pmb p}$ on the neutron Fermi surface,
see Refs.\ \cite{bhy01,gh05} for more details.
%
%
Below in this paper we follow this strategy.

The paper is organized as follows.
In Sec.\ II we formulate a relativistic Hamiltonian describing 
superfluid nucleon-hyperon mixture.
In Sec.\ III we diagonalize this Hamiltonian, 
determine baryon dispersion relations 
in the presence of superfluid currents, 
and calculate the relativistic entrainment matrix $Y_{ik}$.
Section IV presents summary.

Throughout the paper,
unless otherwise stated, 
indices $i$ and $k$ refer to baryons, 
$i$, $k=n$, $p$, $\Lambda$, and $\Sigma$.
We use the system of units
in which the Planck constant $\hbar$,
the speed of light $c$
and the Boltzmann constant $k_{\rm B}$
equal unity, 
$\hbar=c=k_{\rm B}=1$.

\section{The choice of a Hamiltonian}
\label{2}

Let us obtain a model Hamiltonian for a homogeneous system 
of relativistic, degenerate, 
strongly interacting baryons with superfluid currents.
A natural way to obtain such a Hamiltonian is to use 
the framework of Landau Fermi-liquid theory, 
generalized to account for superfluidity 
by Larkin and Migdal \cite{lm63} 
and Leggett \cite{leggett65}.
Following Ref.\ \cite{leggett65},
the Hamiltonian ${\rm H}$ 
describing a superfluid nucleon-hyperon mixture
can be generally presented in the form
%
%
\begin{eqnarray}
{\rm H}- \sum_i \mu_i {\rm N}_i
&=& {\rm H}_{\rm LF} + {\rm H}_{\rm pairing} \,.
\label{hamilt}
\end{eqnarray}
Here, ${\rm N}_{i}$ 
and $\mu_{i}$ 
are, respectively, the number density operator 
and the chemical potential 
for baryon species 
$i=n$, $p$, $\Lambda$, $\Sigma$;
${\rm H}_{\rm LF}$ is the Fermi-liquid Hamiltonian 
for the mixture; 
${\rm H}_{\rm pairing}$ is the pairing Hamiltonian.

If our system were {\it nonrelativistic}, 
${\rm H}_{\rm LF}$ would be given by 
(see, e.g., \cite{leggett65,gh05})
\begin{eqnarray}
{\rm H}_{\rm LF} &=& \sum_{{\pmb p} s i}
\varepsilon_{0}^{(i)} \left({\pmb p} \right) \,
\left( \apc \api - \thp \right)
\nonumber \\
&+& \frac{1}{2}  \sum_{{\pmb p} {\pmb p} '  s s' i k}
f^{i k} \left(\vp, \vps \right)
\left( \apc \api - \thp \right) \left( \apsc \aps - \thps \right).
\label{LF}
\end{eqnarray}
%
Here $\vp$ and $\vps$ are the particle momenta;
$s$ and $s'$ are the spin indices;
$a_{\pmb p}^{(i)} \equiv a_{{\pmb p} s}^{(i)} =
a_{{\pmb p} \uparrow}^{(i)}$ or $a_{{\pmb p} \downarrow}^{(i)}$
is the annihilation operator of a Landau quasiparticle
(not the Bogoliubov excitation!)
of a species $i$ in a state (${\pmb p} s$).
We restrict ourselves to a spin-unpolarized nucleon matter.
This allows us to simplify the notations
by suppressing the spin indices, whenever possible.
Furthermore,
$\thp = \theta \left( \pFa - |\vp| \right)$, where $\theta(x)$
is the step function;
$\varepsilon_{0}^{(i)} \left({\pmb p} \right) = \vFa (|\vp|-\pFa)$,
where $\vFa=\pFa/m^*_i$ and $\pFa$ are, respectively, 
the Fermi-velocity and Fermi-momentum 
with $m^*_i$ being the effective mass.
Finally, $f^{ik}({\pmb p}, {\pmb p'})$
is the spin-averaged
Landau quasiparticle interaction
(we disregard the spin-dependence of this interaction
since it does not affect our results).
In the vicinity of the Fermi surface the arguments of
the function $f^{i k}({\pmb p}, {\pmb p}')$
can be approximately put equal to $p \approx \pFa$ 
and $p' \approx \pFas$, 
while the function itself can be 
expanded into Legendre polynomials $P_l({\rm cos \, \theta})$,
\begin{equation}
f^{i k}({\pmb p}, {\pmb p}') = \sum_l f^{i k}_l \, P_l(\cos \theta),
\label{fik}
\end{equation}
where $\theta$ is the angle between ${\pmb p}$ and ${\pmb p}'$
and $f^{i k}_l$ are the (symmetric) Landau parameters,
$f_l^{ik}=f_l^{ki}$.

%
The Landau theory was extended to relativistic Fermi-liquids 
by Baym and Chin \cite{bc76}.
These authors showed that the formal structure of the theory
remains practically the same as in the nonrelativistic case.
In particular, the expression for the variation 
of the energy density has the same form.
Thus, we assume that the Hamiltonian describing 
the {\it relativistic} mixture is still given by Eq.\ (\ref{LF}).
We emphasize that this is our {\it assumption}; 
it would be interesting to derive this Hamiltonian from the microscopic theory 
in analogy to what was done by Leggett \cite{leggett65} 
for the case of nonrelativistic Fermi-liquid.

The relativistic effects
affect only those properties of the theory 
which are related to transformation of various quantities 
from one frame to another.
For example, as it is demonstrated in Ref.\ \cite{bc76},
the transformation law 
for the Landau quasiparticle interaction $f^{ik}({\pmb p}, {\pmb p'})$
is more complex than that in the nonrelativistic theory.
Also, in the relativistic theory 
the expression for the effective mass $m^*_i$ 
as a function of Landau parameters $f_1^{ik}$ 
should be modified \cite{bc76,gkh09},
\begin{equation}
\frac{\mu_i}{m_i^{\ast}} = 1 - \sum_k \frac{\mu_k G_{ik}}{n_i}.
\label{effmass}
\end{equation}
To reproduce the nonrelativistic result,
one has to replace everywhere in this equation $\mu_i$
with the mass of a free particle $m_i$.
In Eq.\ (\ref{effmass}) 
$n_i$ is the number density of particle species $i$,
\begin{equation}
n_i=\frac{\pFa^3}{3 \pi^2}
\label{numberdensity}
\end{equation}
and the symmetric matrix $G_{ik}$ is defined by
\begin{equation}
G_{ik} \equiv \frac{1}{9 \pi^4} \, \pFa^2 \pFas^2 \, f_1^{ik}.
\label{Gik}
\end{equation}

Let us obtain now an expression 
for the relativistic Hamiltonian ${\rm H}_{\rm pairing}$, 
describing pairing between the Landau quasiparticles.
The pairing in relativistic systems was studied 
in the literature mainly with application to quark matter
(see, e.g., the early review \cite{bl84} 
and a recent review \cite{asrs08} and references therein).
However, there are also several papers that explore 
the importance of relativistic effects 
in terrestrial superconducting materials 
(see, e.g., 
Refs.\ \cite{cg95,cg99a,cg99b,ohsaku02a,ohsaku02b,bertrand05,gb06,beenakker06}).

We start 
with the analysis of a pairing Hamiltonian 
for a mixture of relativistic {\it noninteracting} baryons. 
In other words, we neglect for a moment 
the Landau quasiparticle interaction $f^{ik}({\pmb p}, {\pmb p}')$ 
in Eq.\ (\ref{LF}) for ${\rm H}_{{\rm LF}}$.
As it is demonstrated in Refs.\ \cite{cg95,cg99a,bertrand05,gb06},
in the mean-field approximation
the relativistic analog 
of the well-known non-relativistic Hamiltonian, 
responsible for $^1$S$_0$ pairing of particles, 
has the form
%
\begin{equation}
{\rm H}_{\rm pairing} = \frac{1}{2} \, \sum_i
\int \dd^3 {\pmb r} \, \dd^3 {\pmb r}' \,\, 
\Delta^{(i)}({\pmb r}, {\pmb r}') \,\, 
\left[ 
\overline{\Psi}^{(i)}({\pmb r}) \, \gamma^5 
\, \Psi_{\rm C}^{(i)}({\pmb r}') 
-\frac{1}{2} \, \right. \left\langle  \right| 
\overline{\Psi}^{(i)}({\pmb r}) \, \gamma^5 
\, \Psi_{\rm C}^{(i)}({\pmb r}') 
  \left|   \right\rangle 
\left.  \right]
+ {\rm H.c}.
\label{PairingPSI}
\end{equation}
Here $\Delta^{(i)}({\pmb r}, {\pmb r}')
=\Delta^{(i)}({\pmb r}', {\pmb r})$ 
is the superfluid order parameter;
$\Psi^{(i)}({\pmb r})$ is the relativistic fermion field operator 
and $\Psi^{(i)}_{\rm C}({\pmb r}) 
= i \gamma^2 \gamma^0 \left[ \overline{\Psi}^{(i)}({\pmb r})\right]^{T}$ 
is the charge conjugate field 
(see, e.g., the textbook \cite{bd64} for details).
Furthermore, 
$\overline{\Psi}^{(i)}({\pmb r}) = {\Psi}^{(i) \dagger}({\pmb r}) \gamma^0$; 
$\gamma^5 = i \gamma^0 \gamma^1\gamma^2\gamma^3$, 
where $\gamma^l$ are the Dirac matrices 
($l=0$, $1$, $2$, and $3$). 
The second term in square brackets 
in Eq.\ (\ref{PairingPSI})
is presented to avoid 
double counting in the expression for the system energy. 
Terms of this kind is a typical feature of the mean-field approach.
Similar terms appear in the mean-field formulation 
of the nonrelativistic pairing Hamiltonian 
(see, e.g., a review \cite{gps08}).
%

The Hamiltonian (\ref{PairingPSI}) 
can be simplified if we take into account that 
for a homogeneous system with superfluid currents 
the order parameter $\Delta^{(i)}({\pmb r}, {\pmb r}')$ 
can be written in the form (see, e.g., \cite{bl84})
\begin{equation}
\Delta^{(i)}({\pmb r}, {\pmb r}') = 
\Delta^{(i)}({\pmb r}-{\pmb r}') \,\, {\rm e}^{i {\pmb Q}_i({\pmb r}+{\pmb r}')}=
\sum_{\pmb p} 
\Delta^{(i)}_{\pmb p} \,\, {\rm e}^{i {\pmb p} ({\pmb r}-{\pmb r}')} \,\,
{\rm e}^{i {\pmb Q}_i({\pmb r}+{\pmb r}')}.
\label{order_parameter}
\end{equation}
Here $\Delta^{(i)}_{\pmb p}$ is the 
Fourier component
of the order parameter. 
It is even in ${\pmb p}$,
\begin{equation}
\Delta^{(i)}_{\pmb p} = \Delta^{(i)}_{-{\pmb p}},
\label{gapeven}
\end{equation}
since $\Delta^{(i)}({\pmb r}, {\pmb r}')$ 
is a symmetric function of ${\pmb r}$ and ${\pmb r}'$. 
Furthermore, $2 {\pmb Q}_i$ is the momentum 
of a Cooper pair in condensate.
Below in this paper we always assume that 
$Q_i/\pFas \ll 1$
and restrict ourselves to 
a {\it linear} approximation in $Q_i/\pFas$.

Substituting Eq.\ (\ref{order_parameter}) into Eq.\ (\ref{PairingPSI}),
one gets, after some algebra
[with an accuracy up to quadratic terms in $Q_i/\pFas$],
\begin{eqnarray}
{\rm H}_{\rm pairing} &=& -\sum_{\pmb p i} 
\Delta^{(i)}_{\pmb p}
\left[ 
a_{{\pmb p} +{\pmb Q}_i \uparrow}^{(i) \dagger} \,
a_{-{\pmb p}+{\pmb Q}_i \downarrow}^{(i) \dagger}
-\frac{1}{2} \, \right. \left\langle  \right| 
a_{{\pmb p} +{\pmb Q}_i \uparrow}^{(i) \dagger} \,
a_{-{\pmb p}+{\pmb Q}_i \downarrow}^{(i) \dagger}
  \left|   \right\rangle 
\nonumber\\
 && +c_{{\pmb p} +{\pmb Q}_i \uparrow}^{(i) \dagger} \,
c_{-{\pmb p}+{\pmb Q}_i \downarrow}^{(i) \dagger}
 -\frac{1}{2} \, 
 \left\langle  \right| 
c_{{\pmb p} +{\pmb Q}_i \uparrow}^{(i) \dagger} \,
c_{-{\pmb p}+{\pmb Q}_i \downarrow}^{(i) \dagger}
\left|   \right\rangle 
\left.
\right] + {\rm H. c.}
\label{elpos}
\end{eqnarray}
The analogous derivation of a Hamiltonian for a homogeneous system
is explained in detail in Ref.\ \cite{bertrand05} 
for the case of ${\pmb Q}_i=0$.
In Eq.\ (\ref{elpos}) $c_{\pmb p}^{(i) \dagger}$ 
is the creation operator for {\it antibaryons} of a species $i$.
Generally, because ${\rm H}_{\rm pairing}$ 
does not contain `interference' terms 
(e.g, of the form $a_{{\pmb p} +{\pmb Q}_i \uparrow}^{(i) \dagger} 
c_{{\pmb p} +{\pmb Q}_i \downarrow}^{(i) \dagger}$), 
one may diagonalize the Hamiltonian
${\rm H}_{\rm LF}+ {\rm H}_{\rm pairing}$ 
by simply performing the standard Bogoliubov transformation 
{\it separately} for baryons and antibaryons
(this procedure is very clearly described 
in Ref.\ \cite{bertrand05};  
in this case the species index $i$ in Eq.\ (\ref{LF}) 
for ${\rm H}_{\rm LF}$
should also run over antibaryons).
However, since the conditions in the neutron-star cores are such that
the population of antibaryons is negligible,
one can neglect them and rewrite ${\rm H}_{\rm pairing}$
in its final form
\begin{equation}
{\rm H}_{\rm pairing} =- \sum_{\pmb p i} 
\Delta^{(i)}_{\pmb p} \,\,
\left[
a_{{\pmb p} +{\pmb Q}_i \uparrow}^{(i) \dagger} \,
a_{-{\pmb p}+{\pmb Q}_i \downarrow}^{(i) \dagger}
-\frac{1}{2} \, \right. \left\langle  \right| 
a_{{\pmb p} +{\pmb Q}_i \uparrow}^{(i) \dagger} \,
a_{-{\pmb p}+{\pmb Q}_i \downarrow}^{(i) \dagger}
  \left|   \right\rangle 
   \left. \right]
+ {\rm H. c.}
\label{final_pairing}
\end{equation}
%
%
We see that the Hamiltonian ${\rm H}_{\rm pairing}$ for
a homogeneous system of noninteracting relativistic baryons 
is given by the same equation as in the nonrelativistic limit.
Moreover, it coincides \cite{leggett65, gh05} 
with the pairing Hamiltonian 
describing {\it strongly interacting} nonrelativistic Fermi-liquid
[the operators in Eq.\ (\ref{final_pairing}) 
refer then to Landau quasiparticles rather than to real particles].
Thus, it is reasonable to assume, 
that ${\rm H}_{\rm pairing}$ 
for interacting relativistic baryon mixture
is also given by Eq.\ (\ref{final_pairing}).
All calculations below are made under such assumption.

In the Appendix we give further arguments 
supporting the form of the expression (\ref{final_pairing}) 
for ${\rm H}_{\rm pairing}$.
Namely, 
we apply the consideration given in this section
to a specific case of relativistic mean-field model
in which baryon interactions are mediated 
by various types of meson fields.
We show that for this model the homogeneous Hamiltonian (\ref{final_pairing}) 
can be directly calculated from the general expression (\ref{PairingPSI}).

In view of Eq.\ (\ref{gapeven}), in the linear approximation
the superfluid order parameter $\Delta^{(i)}_{\pmb p}$,
entering Eq.\ (\ref{final_pairing}),
is the same function of ${\pmb p}$
as in the system without superfluid currents. 
It can be chosen {\it real}, 
just as for the system with ${\pmb Q}_i=0$ (e.g., \cite{lp80}).
In this case $\Delta^{(i)}_{\pmb p}$ 
is the energy gap in the dispersion relation
for Bogoliubov quasiparticles (see below).

\section{The relativistic entrainment matrix at finite $T$}

\subsection{General equations}
\label{22}

It follows from Eqs.\ (\ref{LF}) and (\ref{final_pairing})
that the relativistic expressions 
for ${\rm H}_{\rm LF}$ and ${\rm H}_{\rm pairing}$ 
are essentially the same as in the nonrelativistic case.
Correspondingly, 
the further consideration is
quite similar to that in Ref.\ \cite{gh05}.
Thus, here we only briefly sketch the main results, 
referring the reader to Ref.\ \cite{gh05} for more details.

Introducing 
the Bogoliubov operators $b_{{\pmb p} s}^{(i)}$
\begin{eqnarray}
a_{{\pmb p} + {\pmb Q}_{i} \uparrow}^{(i)}
&=& \uup b_{{\pmb p} + {\pmb Q}_{i} \uparrow}^{(i)}
+ \vvp b_{-{\pmb p} + {\pmb Q}_{i} \downarrow}^{(i) \dagger},
\label{bpQ1} \\
a_{{\pmb p} + {\pmb Q}_{i} \downarrow}^{(i)}
&=& \uup b_{{\pmb p} + {\pmb Q}_{i} \downarrow}^{(i)}
- \vvp b_{-{\pmb p} + {\pmb Q}_{i} \uparrow}^{(i) \dagger},
\label{bpQ2}
\end{eqnarray}
where $\uup$ and $\vvp$ 
are even functions of $\vp$ related by the normalization condition,
\begin{equation}
u_{\pmb p}^{(i)2} + v_{\pmb p}^{(i)2} = 1,
\label{normal}
\end{equation}
one obtains for the system energy density $E$
\begin{gather}
E- \sum_i \mu_i n_i =
\sum_{{\pmb p} s i}
\varepsilon_{0}^{(i)}\left({\pmb p}+{\pmb Q}_{i} \right) \,
\left( \mN_{{\pmb p}+ {\pmb Q}_{i}}^{(i)} - \thpQ \right)
\nonumber \\
+ \frac{1}{2}  \sum_{{\pmb p}{\pmb p}'  ss' ik}
f^{i k} \left(\vp + \vQa, \vps + {\pmb Q}_{k} \right)
\left( \mN_{{\pmb p}+ {\pmb Q}_{i}}^{(i)} - \thpQ \right)
\left( \mN_{{\pmb p}'+ {\pmb Q}_{k}}^{(k)} - \thpsQ \right)
\nonumber \\
- \sum_{{\pmb p} i}
\Delta_{\pmb p}^{(i)} \,\,
\uup \vvp \, 
\left(  1- \mF_{{\pmb p}+ {\pmb Q}_{i}}^{(i)}
- \mF_{-{\pmb p}+ {\pmb Q}_{i}}^{(i)}\right).
\label{energy}
\end{gather}
Here $\mN_{{\pmb p}+ {\pmb Q}_{i}}^{(i)}$ and
$\mF_{{\pmb p}+ {\pmb Q}_{i}}^{(i)}$
are the distribution functions for Landau quasiparticles and Bogoliubov excitations
with momentum (${\pmb p}+ {\pmb Q}_{i}$), respectively,
\begin{eqnarray}
\mN_{{\pmb p}+ {\pmb Q}_{i}}^{(i)} &=& \,\,
\langle | a_{{\pmb p} +{\pmb Q}_{i} \uparrow}^{(i) \dagger}
a_{{\pmb p} +{\pmb Q}_{i} \uparrow}^{(i)} |\rangle
\,\, = \,\,
\langle | a_{{\pmb p} +{\pmb Q}_{i}\downarrow}^{(i) \dagger}
a_{{\pmb p}+{\pmb Q}_{i} \downarrow}^{(i)} |\rangle
\nonumber \\
&=& v_{\pmb p}^{(i) \, 2} +
 u_{\pmb p}^{(i) \, 2} \, \mF_{{\pmb p}+ {\pmb Q}_{i}}^{(i)}
- v_{\pmb p}^{(i) \, 2} \, \mF_{-{\pmb p}+ {\pmb Q}_{i}}^{(i)},
\label{np1} \\
\mF_{{\pmb p}+ {\pmb Q}_{i}}^{(i)} &=&
\langle | b_{{\pmb p} +{\pmb Q}_{i} \uparrow}^{(i) \dagger}
b_{{\pmb p} +{\pmb Q}_{i} \uparrow}^{(i)} |\rangle
\,\, = \,\,
\langle | b_{{\pmb p} +{\pmb Q}_{i}\downarrow}^{(i) \dagger}
b_{{\pmb p}+{\pmb Q}_{i} \downarrow}^{(i)} | \rangle.
\label{fp1}
\end{eqnarray}
Notice that, since the parameters $\uup$ and $\vvp$ are 
even functions of ${\pmb p}$, they do not depend on ${\pmb Q}_i$ 
in the linear approximation.

The entropy of the system is given by the standard 
combinatorial expression,
\begin{equation}
S = - \sum_{{\pmb p} s i}
\left[ \left( 1- \mF_{{\pmb p}+ {\pmb Q}_{i}}^{(i)} \right) 
\ln \left( 1-\mF_{{\pmb p}+ {\pmb Q}_{i}}^{(i)} \right) 
+ \mF_{{\pmb p}+ {\pmb Q}_{i}}^{(i)} 
\ln \mF_{{\pmb p}+ {\pmb Q}_{i}}^{(i)}
\right].
\label{entropy}
\end{equation}
Using Eqs.\ (\ref{normal})--(\ref{entropy}) 
and minimizing the thermodynamic potential
$F = E-\sum_i \mu_i n_i -T S$ with respect
to $\mF_{{\pmb p}+ {\pmb Q}_{i}}^{(i)}$ and $\uup$, one gets
\begin{eqnarray}
\mF_{{\pmb p}+ {\pmb Q}_{i}}^{(i)} &=&
\frac{1}{ 1 + {\rm e}^{\mathfrak{E}_{{\pmb p}+ {\pmb Q}_{i}}^{(i)}/T}}
\,,
\label{Fp} \\
\mathfrak{E}_{{\pmb p}+ {\pmb Q}_{i}}^{(i)}
&=& \frac{1}{2} \, \left( H_{{\pmb p}+ {\pmb Q}_{i}}^{(i)}
-H_{-{\pmb p}+ {\pmb Q}_{i}}^{(i)}
\right) \, +
\sqrt{\, \frac{1}{4} \, \left( H_{{\pmb p}+ {\pmb Q}_{i}}^{(i)}
+H_{-{\pmb p}+ {\pmb Q}_{i}}^{(i)}\right)^2
+ \Delta_{\pmb p}^{(i) 2} }\, ,
\label{distribution1} \\
u_{{\pmb p}}^{(i) \, 2} &=& \frac{1}{2} \,
\left( 1 + { H_{{\pmb p}+ {\pmb Q}_{i}}^{(i)}
+H_{-{\pmb p}+ {\pmb Q}_{i}}^{(i)}
\over 2 \mathfrak{E}_{{\pmb p}+ {\pmb Q}_{i}}^{(i)}
+ H_{-{\pmb p}+ {\pmb Q}_{i}}^{(i)}
- H_{{\pmb p}+ {\pmb Q}_{i}}^{(i)}} \right) \,.
\label{upQ}
\end{eqnarray}
Here $\mathfrak{E}_{{\pmb p}+ {\pmb Q}_{i}}^{(i)}$
is the energy of a Bogoliubov excitation of a species $i$ with momentum 
${\pmb p}+{\pmb Q}_i$; 
$H_{{\pmb p}+{\pmb Q}_i}^{(i)}$ is the energy of a Landau quasiparticle 
in normal (nonsuperfluid) matter,
\begin{equation}
H_{{\pmb p}+ {\pmb Q}_{i}}^{(i)} =
\varepsilon_0^{(i)}  \left({\pmb p}+{\pmb Q}_{i} \right)
+ \sum_{{\pmb p}' s' k}
f^{i k} \left(\vp+\vQa, \vps+{\pmb Q}_{k} \right) \,
\left(  \mN_{{\pmb p}' + {\pmb Q}_{k}}^{(k)} - \thpsQ \right).
\label{localenergy2}
\end{equation}
Since $Q_i \ll \pFas$, it can be expanded
in powers of ${\pmb Q}_k$,
\begin{equation}
H_{{\pmb p}+{\pmb Q}_{i}}^{(i)} = \varepsilon^{(i)}({\pmb p}) +
\Delta H_{\pmb p}^{(i)},
\label{expand1}
\end{equation}
where $\varepsilon^{(i)}({\pmb p}) \approx \vFa (|\vp|-\pFa)$ 
is the quasiparticle energy in the absence of superfluid currents and 
$\Delta H_{\pmb p}^{(i)}$ is a small current-dependent term.
In the linear approximation $\Delta H_{\pmb p}^{(i)}$ 
can be generally written as
\begin{equation}
\Delta H_{\pmb p}^{(i)} = \sum_{k} \, \gamma_{ik} \,
\,\, \vp \, {\pmb Q}_{k},
\label{dH}
\end{equation}
where $\gamma_{ik}$ is a $4 \times 4$ matrix to be determined below
(notice, that the definition of $\gamma_{ik}$ 
differs by a factor of $m_k$ from that used in Ref.\ \cite{gh05}).
Substituting Eqs.\ (\ref{distribution1}) 
and (\ref{expand1}) into Eq.\ (\ref{upQ})
one verifies that $\uup$ (and $\vvp$) 
is indeed independent of ${\pmb Q}_{k}$
in the linear approximation.

The distribution function $\mN_{{\pmb p}+ {\pmb Q}_{i}}^{(i)}$
can also be expanded in powers of ${\pmb Q}_{i}$. 
Using Eqs.\ (\ref{np1}) and (\ref{Fp})--(\ref{upQ}), one obtains
\begin{equation}
\mN_{{\pmb p}+{\pmb Q}_{i}}^{(i)} =
\np + {\partial \fp \over \partial 
E_{\pmb p}^{(i)}} \,\, \Delta H_{\pmb p}^{(i)}.
\label{np2}
\end{equation}
Here 
$E_{\pmb p}^{(i)}$, $\fp$, and $\np$
denote, respectively, the quantities 
$\mathfrak{E}_{{\pmb p}+ {\pmb Q}_{i}}^{(i)}$,
$\mF_{{\pmb p}+{\pmb Q}_{i}}^{(i)}$, 
and $\mN_{{\pmb p}+{\pmb Q}_{i}}^{(i)}$,  
in the absence of superfluid currents.
They are given by the following well-known expressions:
\begin{eqnarray}
E_{\pmb p}^{(i)} &=& \sqrt{\varepsilon^{(i) 2} \left({\pmb p} \right)
+ \Delta_{\pmb p}^{(i) 2}},
\label{Ep}\\
\fp &=& \frac{1}{1 + {\rm e}^{E_{\pmb p}^{(i)}/T}}, \quad
 \label{fp}\\
 \np &=& v_{\pmb p}^{(i) \, 2} +
\left( u_{\pmb p}^{(i) \, 2} - v_{\pmb p}^{(i) \, 2}\right) \, \fp.
\label{np}
\end{eqnarray}
%

\subsection{Calculation of the matrix $\gamma_{ik}$}

Let us calculate the matrix $\gamma_{ik}$. 
Substituting Eqs.\ (\ref{expand1}) and (\ref{np2})
into Eq.\ (\ref{localenergy2}), 
one obtains, with the accuracy to the terms linear in ${\pmb Q}_{i}$,
\begin{equation}
\Delta H_{\pmb p}^{(i)} = { {\pmb p \, {\pmb Q}_{i}} \over
m_i^\ast}
+ \sum_{{\pmb p}' s' k} \,
f^{i k} \left( {\pmb p}, {\pmb p}' \right) \,
\left\{
{\partial \mathfrak{f}_{{\pmb p}'}^{(k)} \over \partial E_{{\pmb p}'}^{(k)} }
\,\,
\Delta H_{{\pmb p}'}^{(k)} -
{\partial \theta_{{\pmb p}'}^{(k)}  \over \partial {\pmb p}'} \, {\pmb
Q}_{k}
\right\}.
\label{Eq}
\end{equation}
The functions in curly brackets have sharp maximum 
near the Fermi surface of particle species $k$
(at $p' \sim \pFas$), 
so that the integrals in Eq.\ (\ref{Eq}) can be approximately calculated
and presented in the form
\begin{eqnarray}
\sum_{{\pmb p}' s'} \,
f^{i k} \left( {\pmb p}, {\pmb p}' \right) \,
{\partial \mathfrak{f}_{{\pmb p}'}^{(k)} \over \partial E_{{\pmb p}'}^{(k)} }
\,\,
\Delta H_{{\pmb p}'}^{(k)} &=&  - \frac{G_{ik}}{n_i} \,\, m_k^{\ast} \, \Phi_{k} \,
\Delta H_{{\pmb p}}^{(k)} \, ,
\label{intF} \\
\sum_{{\pmb p}' s'} \,
f^{i k} \left( {\pmb p}, {\pmb p}' \right) \,
{\partial \theta_{{\pmb p}'}^{(k)}  \over \partial {\pmb p}'} \, {\pmb
Q}_{k}
&=& - \, \frac{G_{ik}}{n_i} \,\, {\pmb p} \, {\pmb Q}_{k}.
\label{intQ}
\end{eqnarray}
Here the matrix $G_{ik}$ is defined in Eq.\ (\ref{Gik}) 
while the function $\Phi_{i}$ is given by
\begin{equation}
\Phi_{i} = - 
\frac{\pi^2}{m_i^\ast \pFa}
\, \sum_{{\pmb p} s} \,
{\partial \fp \over \partial E_{\pmb p}^{(i)}}.
\label{N0}
\end{equation}
It changes from $\Phi_i = 0$ at $T = 0$ 
to $\Phi_i = 1$ at $T \geq T_{ci}$
(we remind that $T_{ci}$ is the critical temperature 
for transition of particle species $i$ 
to the superfluid state).
Assuming that the energy gap $\Delta_{\pmb p}^{(i)}$ does not depend 
on the momentum ${\pmb p}$, 
this function was calculated numerically 
and fitted in Ref.\ \cite{gy95}.
The fitting formula and further details concerning $\Phi_i$ 
can also be found in Ref.\ \cite{gh05}.

Using Eqs.\ (\ref{dH}), (\ref{intF}), and (\ref{intQ}) 
and equating prefactors in front of the same $Q_k$ in Eq.\ (\ref{Eq}),
one obtains the following system of $4 \times 4$ 
linear equations for the matrix $\gamma_{ik}$:
\begin{equation}
\gamma_{ik} = \frac{\delta_{ik}}{m_{i}^{\ast}} + \frac{G_{ik}}{n_i}
-\sum_l \frac{G_{il}}{n_i} \,\, m_l^{\ast} \,\,\Phi_l \,\, \gamma_{lk}.
\label{gamma}
\end{equation}
Notice that this matrix is {\it not} symmetric, $\gamma_{ik}\neq \gamma_{ki}$.
In Eq.\ (\ref{gamma}) for each $k$ 
we have four equations (with $i=n$, $p$, $\Lambda$, and $\Sigma$).
These equations can be decoupled from the whole system and solved separately.
Thus, there are in fact four independent systems of four linear equations.
Their solutions can be immediately found in a number of limiting cases:

($i$) For very low temperatures, when all $\Phi_i=0$,
\begin{equation}
\gamma_{ik} = \frac{\delta_{ik}}{m_{i}^{\ast}} + \frac{G_{ik}}{n_i}.
\label{lowT}
\end{equation}

($ii$) For nonsuperfluid matter, when all $\Phi_i=1$,
\begin{equation}
\gamma_{ik} = \frac{\delta_{ik}}{m_{i}^{\ast}}.
\label{highT}
\end{equation}
%

($iii$) If the species in the mixture do not interact with each other, 
that is $f^{ik}_l=0$ for $i\neq k$, 
one has [see Eqs.\ (\ref{effmass}) and (\ref{Gik})]
\begin{equation}
\gamma_{ik} = \frac{n_i \,\,  \delta_{ik}}
{\mu_i \left( n_i + G_{ii} \, m_i^{\ast} \,\Phi_i  \right)}.
\label{gamma1}
\end{equation}

($iv$) If $\Phi_i = 0$ for all species except for
the species $j$ (i.e. $\Phi_j \neq 0$), then
\begin{equation}
\gamma_{ik} = \frac{\delta_{ik}}{m_{i}^{\ast}} + \frac{G_{ik}}{n_i}
-\frac{G_{ij} \, (G_{jk} \, m_j^{\ast} + n_j \, \delta_{jk}) \, \Phi_j}
{n_i \, (n_j + G_{jj} \, m_j^{\ast} \, \Phi_j)}.
\label{4limit}
\end{equation}

($v$) Finally, for any nonsuperfluid species $j$ (for which $\Phi_j=1$), 
the matrix elements $\gamma_{ij}$ 
related to this species, are given by the same 
Eq.\ (\ref{highT}) as in the case of 
a completely nonsuperfluid mixture,
\begin{equation}
\gamma_{ij} 
=\frac{\delta_{ij}}{m_{i}^{\ast}}.
\label{high2}
\end{equation}

The general solution for the matrix $\gamma_{ik}$ in case of four particle species 
is rather lengthy, but can be easily obtained from Eq.\ (\ref{gamma}) 
and, for example, incorporated into a computer code.
Here we present the solution for an important case of 
a mixture of two particle species (e.g., neutrons and protons)
\begin{eqnarray}
\gamma_{ii}&=&\frac{\left(  n_i + G_{ii} \, m_i^{\ast} \right)
\left(  n_k + G_{kk}\, m_k^{\ast}\, \Phi_k \right) 
- G_{ik}^2 \,m_i^{\ast}\, m_k^{\ast} \,\Phi_k}
{m_i^{\ast}\,\, S}, 
\label{solution1}\\
\gamma_{ik} &=& \frac{G_{ik} \, n_k\, (1-\Phi_k)}{S},
\label{solution2}
\end{eqnarray}
where
\begin{equation}
S=\left(  n_i + G_{ii} \, m_i^{\ast} \, \Phi_i \right)
\left(  n_k + G_{kk}\, m_k^{\ast} \, \Phi_k \right) 
- G_{ik}^2 \, m_i^{\ast} \, m_k^{\ast} \, \Phi_i \, \Phi_k. 
\label{S}
\end{equation}
In Eqs.\ (\ref{solution1})--(\ref{S}) 
indices $i$ and $k$ refer to different particle species, $i \neq k$. 
The presented solution agrees with that given in Ref.\ \cite{gh05}
for the nonrelativistic case (we remind the reader 
that the definition of $\gamma_{ik}$ 
in Ref.\ \cite{gh05} slightly differs from ours, 
see the comment to Eq.\ (\ref{dH})).
However, in contrast to Ref.\ \cite{gh05}, 
in Eqs.\ (\ref{solution1})--(\ref{S})
one should use the {\it relativistic} expression 
for the effective masses given by Eq.\ (\ref{effmass}).

\subsection{Calculation of the matrix $Y_{ik}$}

To find the relativistic entrainment matrix $Y_{ik}$ let us calculate 
the particle current density ${\pmb j}_i$. 
For that, we make use of the fact that, 
as was emphasized by Leggett \cite{leggett65}, 
the expression for ${\pmb j}_i$ has the same form as 
in the case of usual nonsuperfluid Fermi-liquid 
(see also Refs.\ \cite{bjk96,cj03,gh05,ch06,gkh09}). 
In other words, 
\begin{equation}
{\pmb j}_i = \sum_{{\pmb p} s} 
\frac{\partial H^{(i)}_{{\pmb p}+{\pmb Q}_i}}{\partial {\pmb p}} \,\,
\mathcal{N}^{(i)}_{{\pmb p}+{\pmb Q}_i}.
\label{ji}
\end{equation}
Substituting expressions (\ref{expand1}) and (\ref{np2}) 
into this equation 
and performing a simple integration, 
one obtains Eq.\ (\ref{jnn}) with
\begin{equation}
Y_{ik} = n_i \,\, \gamma_{ik} \,\, (1-\Phi_i).
\label{Yik}
\end{equation}\
One may check that this matrix is symmetric, $Y_{ik}=Y_{ki}$.
At zero temperature this equation can be simplified 
with the help of Eq.\ (\ref{lowT}), 
so that one reproduces the result of GKH09,
\begin{equation}
Y_{ik} = \frac{n_i}{m_i^{\ast}} \, \delta_{ik} + G_{ik}.
\label{T=0}
\end{equation}
Notice that, as follows from Eq.\ (\ref{effmass}), 
in this limit the matrix $Y_{ik}$ satisfies the sum rule (see GKH09),
\begin{equation}
\sum_k \mu_k \, Y_{ik} = n_i.
\label{sumrule}
\end{equation}
%

In the nonrelativistic limit the matrix $Y_{ik}$ 
is related to the nonrelativistic entrainment matrix $\rho_{ik}$
by the equation (see, e.g., \cite{ga06})
\begin{equation}
\rho_{ik} = m_i \, m_k \, Y_{ik}.
\label{nonrel}
\end{equation}
Using this equation and Eq.\ (\ref{Yik}),
one may reproduce the various limiting formulas for $\rho_{ik}$, 
presented in the literature \cite{leggett65,bjk96,gh05}.

Our results are illustrated in the Fig.\ 1.
The normalized symmetric matrix $Y_{ik}/Y$ is shown 
as a function of temperature $T$ for 
the baryon number density $n_b=3 n_0=0.48$ fm$^{-3}$.
To plot the figure we employed
the {\it third} equation of state of Glendenning \cite{glendenning85}.
The Landau parameters $f_1^{ik}$ 
of nucleon-hyperon matter were calculated 
for this equation of state in GKH09.
The normalization constant $Y$ is taken to be 
$Y=3n_0/\mu_n(3 n_0)=2.48 \times 10^{41}$ erg$^{-1}$ cm$^{-3}$,
where $n_0=0.16$ fm$^{-3}$ is the normal nuclear density and
$\mu_{n}(3 n_0) = 1.94 \times 10^{-3}$ erg 
is the neutron chemical potential at $n_b =3 n_0$.
We choose the baryon critical temperatures $T_{ci}$ 
($i=n$, $p$, $\Lambda$, $\Sigma$) equal to: 
$T_{cn}=5 \times 10^8$ K, $T_{cp}=2 \times 10^9$ K, 
$T_{c \Lambda}=3 \times 10^9$ K, and
$T_{c \Sigma}= 6 \times 10^9$ K.
%
\begin{figure}[t]
\setlength{\unitlength}{1mm}
\leavevmode
\hskip  0mm
\includegraphics[width=150mm,bb=15  300  550  550,clip]{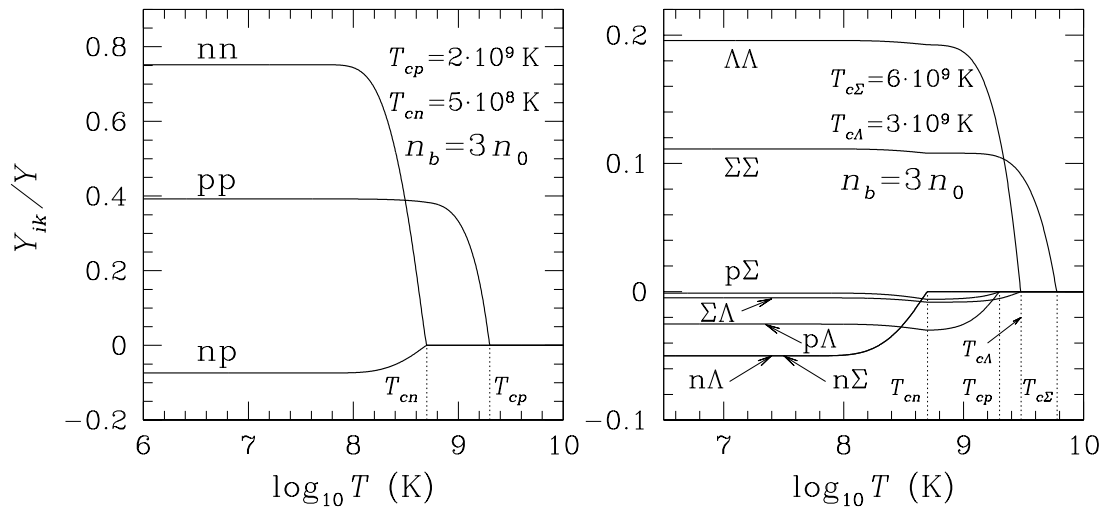}
\caption
{
Normalized symmetric matrix $Y_{ik}/Y$ 
as a function of $T$ for the third equation 
of state of Glendenning \cite{glendenning85}.
The normalization constant $Y=3 n_0/\mu_{n}(3 n_0) 
= 2.48 \times 10^{41}$ erg$^{-1}$ cm$^{-3}$.
Solid lines show the elements of the matrix $Y_{ik}/Y$;
each curve is marked by the corresponding symbol $ik$
($i, k = n$, $p$, $\Lambda$, $\Sigma$).
Vertical dotted lines indicate baryon critical temperatures.}
\label{1fig}   
\end{figure}

On the left panel we plot the `nucleon' matrix elements; 
the `hyperon' matrix elements are shown on the right panel. 
In agreement with Eq.\ (\ref{Yik}), the matrix elements, 
involving a nonsuperfluid species, vanish.
For instance, if neutrons are normal (i.e., $T>T_{cn}$), 
then $\Phi_n=1$ and $Y_{ni}=Y_{in}=0$.
At $T \la 10^8$ K
all the matrix elements approach 
their asymptotic zero-temperature values [see Eq.\ (\ref{T=0})].
As follows from the figure, the nondiagonal matrix elements
are essentially smaller than the diagonal ones and are all negative 
for the chosen equation of state.

\section{Summary}

The present paper is a continuation of GKH09, 
where the relativistic entrainment matrix $Y_{ik}$ 
of nucleon-hyperon mixture
was calculated for the case of zero temperature.
Here we extend the results of GKH09 to finite temperatures.
For that, we employ the relativistic Landau Fermi-liquid theory \cite{bc76},
generalized to allow for superfluidity of baryons \cite{lm63,leggett65}.

We demonstrate that, as in the case of nonrelativistic 
neutron-proton mixture \cite{gh05},
the matrix $Y_{ik}$ is expressed through the Landau parameters $f_1^{ik}$
and the function of temperature $\Phi_i$.
The Landau parameters $f_1^{ik}$ for the relativistic nucleon-hyperon matter 
were derived in GKH09.
The quantity $\Phi_i(T)$ is the universal function of $T$ 
under assumption that the baryon energy gap $\Delta^{(i)}_{\pmb p}$ 
is independent of momentum ${\pmb p}$.
In the latter case the fitting formula for $\Phi_i$ 
can be found in Refs.\ \cite{gy95,gh05}.

Our consideration of $Y_{ik}$ differs from the previous calculations 
of the entrainment matrix, 
available in the neutron-star literature \cite{bjk96,cj03,gh05,ch06}, 
by the following:

($i$) we consider the problem in a fully relativistic framework;

($ii$) we allow for the presence of two hyperon species 
($\Lambda$- and $\Sigma^-$-hyperons), 
in addition to neutrons and protons.

It should be noted that our results can be easily extended to describe 
any number of baryon species (not necessarily four).
The main problem then will be to determine the Landau parameters 
for these species.

The calculated relativistic entrainment matrix $Y_{ik}$ 
is an essential ingredient 
in hydrodynamics of superfluid mixtures \cite{gk08}.
It can be important for studying the pulsations of massive neutron stars
with superfluid nucleon-hyperon cores 
(see Ref.\ \cite{kg09} for an example of such study).
Also, since the matrix $Y_{ik}$ enters the expression (\ref{Fp})
for the equilibrium distribution function of Bogoliubov excitations, 
it can influence various kinetic properties of superfluid baryon matter,
for example, the shear viscosity. 
The related problems will be analyzed in a separate publication.

\section*{Appendix}

The general approach of Sec.\ II can be illustrated 
with a specific example
of a relativistic mean-field model 
in which interactions between baryons 
are mediated by meson fields. 
For definiteness, following GKH09, 
we consider $\sigma$-$\omega$-$\rho$ version of the mean-field 
model with self-interactions of scalar $\sigma$-field.
However, 
our consideration 
remains essentially 
unaffected if one allows for additional meson fields 
(e.g., $\delta$-meson or hidden strangeness 
$\sigma^\ast$ and $\phi$-meson fields).

In GKH09 we reformulated the $\sigma$-$\omega$-$\rho$ mean-field model
in terms of the relativistic Landau Fermi-liquid theory.
In particular, we calculated the Landau quasiparticle interaction function 
$f^{ik}({\pmb p}, {\pmb p}')$. 
Thus, we determined
the Hamiltonian ${\rm H}_{\rm LF}$ 
for this model [see Eq.\ (\ref{LF})].
Let us now turn to the Hamiltonian
${\rm H}_{\rm pairing}$.
It is still given by Eq.\ (\ref{PairingPSI}) with
\begin{eqnarray}
\Psi^{(i)}({\pmb r}) &=& \sum_{{\pmb p} s} \frac{1}{\sqrt{2 \mathcal{E}_i}} 
\left(
a_{{\pmb p}s}^{(i)} \,\, u^{(i)}({\pmb p}, s) \,\, {\rm e}^{i {\pmb p}{\pmb r}}
+ c^{(i)\dagger}_{{\pmb p}s} \,\, \nu^{(i)}({\pmb p}, s) \,\, {\rm e}^{-i {\pmb p}{\pmb r}}
\right),
\label{Psi}\\
\Psi^{(i)}_{\rm C}({\pmb r}) &=& \sum_{{\pmb p} s} 
\frac{i}{\sqrt{2 \mathcal{E}_i}} 
\left(
a_{{\pmb p}s}^{(i) \dagger} \,\, \nu^{(i)}({\pmb p}, s) \,\, 
{\rm e}^{-i {\pmb p}{\pmb r}} +
c^{(i)}_{{\pmb p}s} \,\, u^{(i)}({\pmb p}, s) \,\,
{\rm e}^{i {\pmb p}{\pmb r}}
\right). 
\label{PsiC}
\end{eqnarray}
The bispinors $u^{(i)}({\pmb p}, s)$ and $\nu^{(i)}({\pmb p}, s)$ correspond, 
respectively, to particles and antiparticles.
For the mean-field model they can be written out explicitly 
(see, e.g., Ref.\ \cite{glendenning00}),
\begin{equation}
u^{(i)}({\pmb p}, s) = \binom{\sqrt{\mathcal{E}_i+M^\ast_i} \,\, w_s}
{\sqrt{\mathcal{E}_i-M^\ast_i} \,\, ({\pmb {\rm n}}{\pmb \sigma}) \, w_s}, \quad\quad
\nu^{(i)}({\pmb p}, s) = \binom{\sqrt{\mathcal{E}_i-M^\ast_i} \,\, 
({\pmb {\rm n}}{\pmb \sigma}) \, w_{-s}^{'}}
{\sqrt{\mathcal{E}_i+M^\ast_i} \,\, w_{-s}^{'}}.
\label{spinor1}
\end{equation}
In Eqs.\ (\ref{Psi})--(\ref{spinor1})
$\mathcal{E}_i=\sqrt{{\pmb P}^2 + M_{i}^{\ast 2}}$;
unit vector ${\pmb {\rm n}}$ is directed along ${\pmb P}$;
${\pmb \sigma}=(\sigma_x, \sigma_y, \sigma_z)$ is the vector composed of Pauli matrices;
$w_s$ and $w_s^{'}$ are the spinors 
which are defined as in Ref.\ \cite{blp82} 
(see $\S 23$ of this reference).
Furthermore,
\begin{eqnarray}
{\pmb P} &=& {\pmb p} 
- g_{\omega i} {\pmb \omega} 
- g_{\rho i} I_{3i} {\pmb \rho}_3,
\label{momentum1}\\
M_i^\ast &=& m_i - g_{\sigma i} \sigma,
\label{mass1}
\end{eqnarray}
where ${\pmb \omega}$, ${\pmb \rho}_3$, and $\sigma$ are the meson fields
that are generated by baryon currents and densities; 
$g_{\omega i}$, $g_{\rho i}$, and $g_{\sigma i}$ are the coupling constants;
and $I_{3i}$ is the isospin projection for baryon species $i$
(for more details, see, e.g., GKH09).

Substituting now field operators (\ref{Psi}) and (\ref{PsiC}) 
into the expression (\ref{PairingPSI})
for ${\rm H}_{\rm pairing}$, one verifies, 
that the homogeneous pairing Hamiltonian 
for the mean-field $\sigma$-$\omega$-$\rho$ model
is given by the same expression as was obtained in Sec.\ II 
[see Eq.\ (\ref{elpos}) of this section or Eq.\ (\ref{final_pairing}), 
if we neglect antibaryons].
This particular example presents an additional argument 
supporting 
the conjecture
that the general expression for ${\rm H}_{\rm pairing}$
in the relativistic Landau Fermi-liquid theory,
is also of the form (\ref{final_pairing}).
%

\section*{Acknowledgments}

This research was supported in part
by RFBR (Grants 08-02-00837 and 05-02-22003),
by the Federal Agency for Science and Innovations
(Grant NSh 2600.2008.2), and 
by the Polish MNiSW (Grant N20300632/0450). 
Two of the authors (M.E.G. and E.M.K) acknowledges 
support from the Dynasty Foundation.
Finally, M.E.G. also acknowledges support 
from the RF Presidential Program 
(grant MK-1326.2008.2).


\end{document}